# Laser undulator by laser ponderomotive force

Zhinan Zeng,[1,2] Ruxin Li,[1,2,3] and Zhizhan Xu[1,2,3]

[1]State Key Laboratory of High Field Laser Physics, Shanghai Institute of Optics and Fine Mechanics, Chinese Academy of Sciences, Shanghai 201800, China

[2]IFSA Collaborative Innovation Center, Shanghai Jiao Tong University, Shanghai 200240, China

[3]School of Physical Science and Technology, ShanghaiTech University, Shanghai 200031, China

**Abstract**

The laser-plasma accelerator has attracted great interest for constituting an alternative in the production of the relativistic electron beams of high peak current. But the generated electron beam has poor monochrome and emittance, which make it difficult to produce high brightness radiation. Here we propose a compact flexible laser undulator based on ponderomotive force to constitute a millimeter-sized synchrotron radiation source of X-ray. We demonstrate that it can produce bright radiation with tens of keV energies. This opens a new path to compact synchrotron source with high power laser.

1. Introduction

X-ray is very important for the understanding of the matters, and creates new sciences and technologies. Today, the brightest X-ray beams are generated by free electron laser based on synchrotron radiation (SR) [1]. SR is produced by electrons, which are accelerated to relativistic energies $E = \gamma mc^2$, where $\gamma$ is the Lorentz factor, m is the rest electron mass and c is the light velocity in vacuum. If the relativistic electron ($\gamma \gg 1$) goes through a spatially modulated field, it will oscillate in the field and radiates light at a frequency, which is $2\gamma^2$ times of their own oscillating frequency. In conventional SR sources, the field is produced by a periodic structure of permanent magnets, which is metre-long with a period of 1–10 centimetres for efficient X-ray generation.

In the conventional SR facilities, the relativistic electron beams are typically generated by large-scale accelerator. With the development of the laser technologies, the laser-plasma accelerator (LPA) constitutes an alternative in the production of the relativistic electron beams. In the past few years, the electron beam with peak current above 10kA, low emittance less than 0.1mm and beam energy of 1GeV has been generated [2-11]. With this kind of electron beams and the magnetic undulators, generation of visible light and soft x-ray emission has recently been demonstrated [12-16]. But, the electron beams produced by LPA have big divergence, typically a few milliradian, which will decrease the current density greatly and quickly, even only going through a few periods of conventional magnetic undulator, and strongly affect the brightness of the SR. Although many work can be done to enlarge the election beam size and decrease the divergence, the advantage of this kind of electron beam, high current density, will loss.

To overcome this limitation, many novel, short period wiggler/undulators have been proposed. A kind of short period wiggler/undulator is based on microfabrication [17-25]. Combination of the microstructure and the strong laser pulse can offer the short period wiggler/undulator with sub-millimeter period. Another kind of short period wiggler/undulator is to use the laser period itself, which can offer very short period of laser wavelength. But as we known, the longest laser wavelength is about ten micrometers, it is too short to produce coherent radiation as an undulator for GeV electron beam [26-30].

In the relativistic regime, the interaction of ultraintense laser pulses with free electrons in vacuum has been studied [31]. The averaged equation of motion of the electron in the laboratory frame finally writes as

$$\frac{d\bar{p}}{dt} = -\frac{1}{2m\gamma}\nabla\left(\overline{\left|\frac{eA}{c}\right|^2}\right) \quad (1)$$

where e, m, c, A, γ and p are electron charge, rest electron mass, light velocity in vacuum, laser vector potential in the Coulomb gauge, Lorentz factor and the electron momentum, respectively, wave, and where the overbar denotes an average over the laser period. If the gradient of the electron energy γ can be neglected, we can write the ponderomotive potential as

$$U = \frac{1}{2m\gamma}\left(\overline{\left|\frac{eA}{c}\right|^2}\right) \quad (2)$$

Using the normalized peak amplitude of the laser pulse $a = \frac{eA}{mc^2}$, the ponderomotive potential can be rewritten in the unit of the rest mass energy of the electron as, $U = \frac{a^2 mc^2}{2\gamma} = \frac{a^2}{2\gamma}\cdot[0.511 MeV]$. So, for the electron with γ = 1000, the laser pulse of a = 1 can give the potential barrier of 0.26keV, which means that the electron of 1 mrad divergence angle can be restricted by this potential in the transverse. In the work of Ph. Balcou et al [32], a Raman X-ray free electron laser is proposed and carefully analyzed. It may open perspectives for ultracompact coherent light sources up to the hard x-ray range. In the works of Ph. Balcou et al, the femtosecond or picosecond intense laser is split into two strictly identical parts. These two parts will counter-propagate and interfere to form a standing wave in the transverse in the rest frame of the electron. When the electron goes through this standing wave with a small initial transverse velocity, it will wiggle by the ponderomotive force and produce the Raman radiation. The oscillation frequency is proportional to the electrical strength positively and the Lorentz factor inversely.

In this work, we propose a new kind of laser undulator scheme with configured laser field, which can be realized and controlled with diffraction optics. From eq. (1) we can see, the ponderomotive force is from the spatial gradient of the laser field, so we can design the structure of the laser field in spatial to control the electron motion carefully and arbitrarily.

2. Proposed scheme

In the conventional magnetic undulator of free electron laser, the transverse electron velocity can be written as a sine function $\beta = -\frac{K}{\gamma}\sin(k_u z)$, where K is the undulator parameter, γ is the Lorentz factor and $k_u = 2\pi/\lambda_u$, $\lambda_u$ is the undulator period.

In the electron rest frame, if the laser field can be seen as similar to the structure in Fig. 1(a), the ponderomotive force of the laser field can work as the magnetic field in the undulator in the free electron laser for the electron going through. In the lab frame, this laser field can be look like that shown in Fig. 1(b), which can be produced with wavefront tilting technique. Since the electron velocity and the group velocity of the laser pulse are both close to the light speed in vacuum, the wavefront tilting angle is about 45 degree.

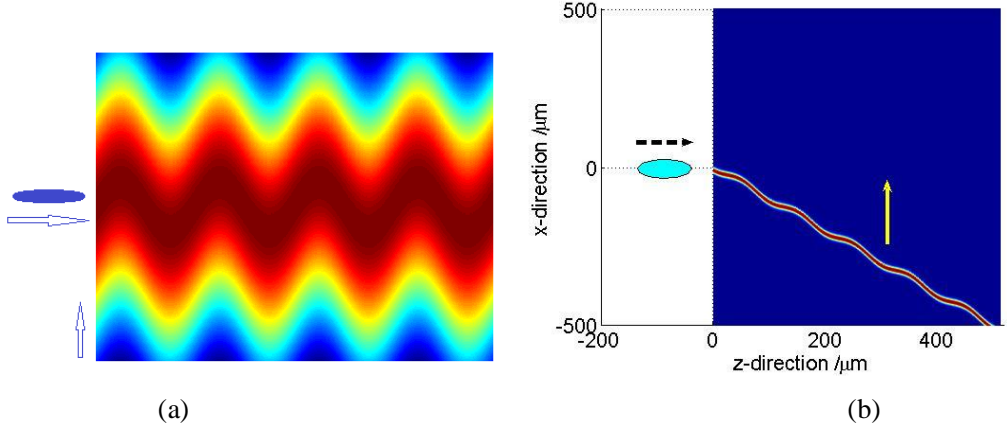

(a)            (b)

Fig. 1 (a) laser field in electron rest frame;    (b) laser field in lab frame;

In Fig. 1(b), the initial velocity of the electron is in z axis. The laser field propagates in x axis. So we can write the laser vector potential as below

$$A = A_0 \exp\left(-\left(t - \frac{x}{c} - \frac{z tg\theta + \Delta \cos k_u z}{c}\right)^2 / \tau^2\right) \cos(\omega t - kx) \quad (3)$$

Where $A_0$, $\omega$, $k$ and $\tau$ are peak vector potential, angular frequency, wave vector, pulse duration of the laser pulse, respective, and c is the light velocity in vacuum. $\theta$ is the wavefront tilting angle. $\Delta$ is the amplitude of wavefront bending and $k_u$ describes the period of the wavefront bending, which can be realized and controlled by diffraction optics technique.

With eq. (3), we can calculate the ponderomotive force using eq. (1) as

$$|A|^2 = \left|A_0 \exp\left(-\left(t - \frac{x}{c} - \frac{z tg\theta + \Delta \cos k_u z}{c}\right)^2 / \tau^2\right) \cos(\omega t - kx)\right|^2$$

$$\bar{I} = \frac{A_0^2}{2} \exp\left(-2\left(t - \frac{x}{c} - \frac{z tg\theta + \Delta \cos k_u z}{c}\right)^2 / \tau^2\right)$$

Then the ponderomotive force in x axis can be written as

$$\bar{F}_x = \frac{e^2}{c^2} \frac{\partial \bar{I}}{\partial x} = \frac{e^2 A_0^2}{2c^2} \exp\left(-\frac{2}{\tau^2}\left(t - \frac{x}{c} - \frac{z tg\theta + \Delta \cos k_u z}{c}\right)^2\right) \frac{4\left(t - \frac{x}{c} - \frac{z tg\theta + \Delta \cos k_u z}{c}\right)}{c\tau^2}$$

(4)

The electron mainly goes in z axis. Simply we can written the position of the electron in z axis as

$$z = c\beta_0 t$$

We can take it into eq. (4) and get

$$\bar{F}_x = \frac{e^2 A_0^2}{2c^2} \exp\left(-2\left(t - \frac{x + c\beta_0 t\,tg\theta}{c} - \frac{\Delta \cos k_u c\beta_0 t}{c}\right)^2 / \tau^2\right) \frac{4\left(t - \frac{x + c\beta_0 t\,tg\theta}{c} - \frac{\Delta \cos k_u c\beta_0 t}{c}\right)}{c\tau^2}$$

In x axis, as the electron mainly moves around x = 0, we can calculate the wavefront tilting angle with

$$t - \frac{c\beta_0 t}{c} tg\theta = 0 \qquad \Rightarrow tg\theta = \frac{1}{\beta_0}$$

If the wavefront of the laser pulse can be tilt in this angle accurately, we can get

$$\bar{F}_x = \frac{e^2 A_0^2}{2c^2} \exp\left(-2\left(-\frac{x}{c} - \frac{\Delta \cos k_u c\beta_0 t}{c}\right)^2 / \tau^2\right) \frac{4\left(-\frac{x}{c} - \frac{\Delta \cos k_u c\beta_0 t}{c}\right)}{c\tau^2} \qquad (5)$$

For the electron precisely located at x = 0, the ponderomotive force can be written as

$$\bar{F}_x = \frac{e^2 A_0^2}{2c^2} \exp\left(-2\left(-\frac{\Delta}{c} \cos k_u c\beta_0 t\right)^2 / \tau^2\right) \frac{4\left(-\frac{\Delta}{c} \cos k_u c\beta_0 t\right)}{c\tau^2}$$

It has very good periodicity. So, under the work of this force, the x axis motion of the electron will also have good periodicity. To demonstrate it, we will give some numerical simulation below.

In the simulation, the 6D phase space of the electron beam is filled with Hammersley sequences as used in the *Genesis* code, but only the motion in the transverse plane (x-y plane) is used in the calculation. The Lorentz factor of the initial electron beam is $\gamma = 1000$. Similar to the electron beam generated in the laser-plasma-driven acceleration, the initial beam size and the emittance of the electron beam are 2μm and $1\times10^{-6}$ m·rad, respectively. Then the electron beam propagates and spreads. After propagating a little distance, the electron beam will be refocused by an ideal focusing system to be about 2.5μm. Fig. 2 shows the electron beam size around the focal point in a 128 periods laser undulator with a period of 100μm.

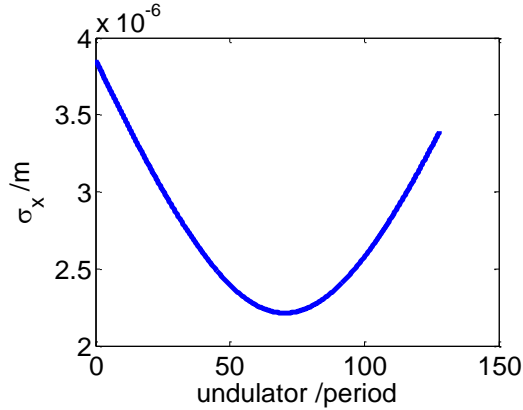

Fig. 2 the electron beam size in x axis ($\sigma_x$) around the focal point.

The ponderomotive force in x axis is written as eq. (5). During the following simulation, the undulator period of 100μm is used. The amplitude of wavefront bending in eq. (4) is 10μm. The duration of the laser pulse is 30fs. In Fig. 3, the electron momentum in x axis is shown when the electron beam goes through the laser undulator.

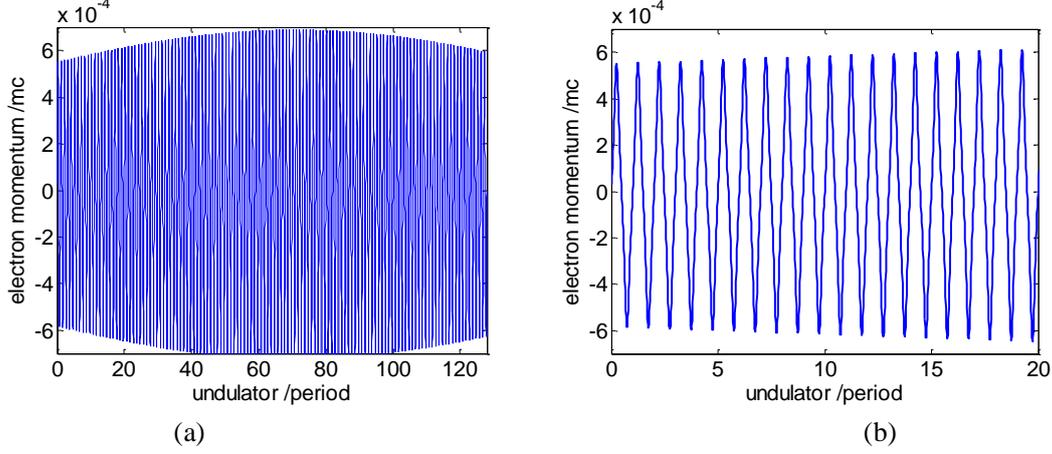

(a)               (b)

Fig. 3 the electron momentum in x axis is shown. The amplitude of the laser electric field is $5\times10^{12}$ V/m, corresponding to the laser intensity of $3.5\times10^{18}$ W/cm$^2$.

Then we can use the following formula to calculate the electron radiation in the far field (Liénard–Wiechert fields),

$$\frac{\partial^2 P}{\partial\Omega\partial\omega} = \frac{e^2}{4\pi\varepsilon_0 \cdot 4\pi^2 c}\left|\int_{-\infty}^{+\infty} dt \frac{n\times\left[(n-\vec{\beta})\times\dot{\vec{\beta}}\right]}{(1-n\cdot\vec{\beta})^2} e^{i\omega(t-n\cdot\vec{r}/c)}\right|^2$$

where n is the unit vector denoting the radiation direction, β is the normalized electron velocity, ω is the radiation frequency, r is the position of the electron, c is the light velocity in vacuum, ε0 is vacuum permittivity. In the simulation, the electron beam is assumed as 50pC charge and 2μm length. The electron density distribution on axis is uniform and the electron radiation of different position on axis is summed incoherent. Fig. 4 is the simulation results, in which we show the electron radiation spectra under two different laser electric field strength, where $\omega_0 = ck_u \cdot 2\gamma^2$, about 24.8keV here. When the peak electric strength can be 5e12 V/m, the peak brightness can be the $6.3\times10^{25}$ photons s$^{-1}$ mm$^{-2}$ mrad$^{-2}$ per 0.1%bandwidth. The FWHM of the bandwidth is about 0.11keV. Fig. 4b shows the peak brightness under different laser electric field from 1e12 V/m to 5e12 V/m, where the peak brightness increases as $E_0^4$ ($E_0$ is the peak electric strength.).

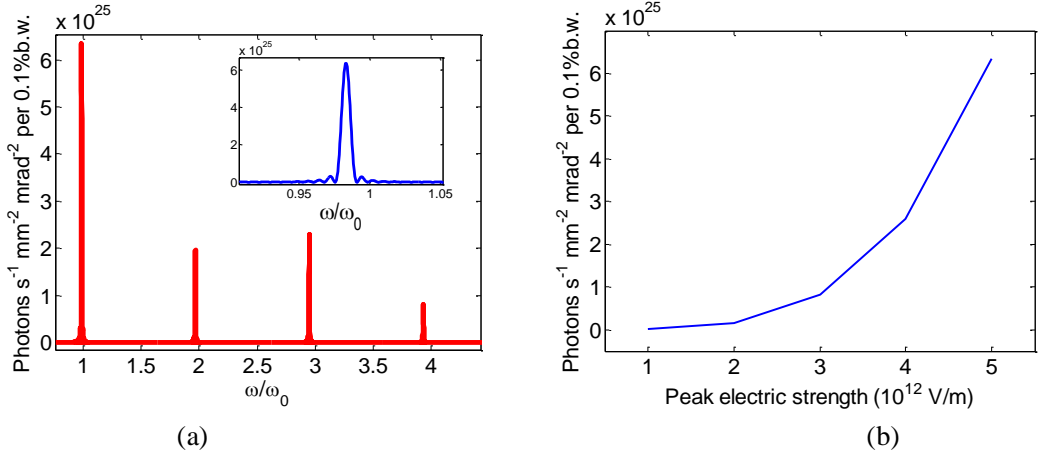

(a)               (b)

Fig. 4 the radiation spectra on axis for different laser electric strengths

But as we known, the brightness of soft x-ray free electron laser (FEL) usually can be $10^{33}$ photons $s^{-1}mm^{-2}$ $mrad^{-2}$ per 0.1%bandwidth, can we increase the laser intensity of this kind of laser undulator to achieve it? From Fig. 3(b), we can know that the undulator parameter is about $6\times10^{-4}$. If we increase it to be 0.6, close to the undulator parameter of the conventional FEL, the laser intensity need to be about $3.5\times10^{21}$ $W/cm^2$, lower than the previous scheme [22]. To demonstrate it, we use the *genesis* code to perform the simulation. During the simulation, the laser undulator is simplified as a 100 μm period conventional FEL. With the increase of the undulator parameter, the wavelength of the radiation will also increase, e.g. 0.069nm (about 18keV) for 0.6 undulator parameter. But we still find that the emission power can increase as $E_0^4$ approximately, up to the brightness of several times of $10^{31}$ photons $s^{-1}mm^{-2}$ $mrad^{-2}$ per 0.1%bandwidth for 0.6 undulator parameter. If we decrease the electron energy to γ = 100 to produce the soft x-ray radiation, the wavelength of the emission will be 3.46nm for 0.6 undulator parameter and the emission power looks like squarely increasing with *nwig* number in *genesis* code, which means the emission can be coherent.

In summary, we have presented a new laser undulator that can be achieved with high power femtosecond laser, which can drastically reduce the size and the cost. The parameter of the laser undulator can be flexibly designed and good control, e.g. different undulator period can be produced with different focal length. We have also shown that this laser undulator can produce bright quasi-monoenergetic X-ray beams with a peak brightness of the order of $10^{25}$ photons $s^{-1}mm^{-2}mrad^{-2}$ per 0.1% bandwidth. If the laser undulator is carefully controlled, the free electron laser may be generated with high quality electron beam. We anticipate that this kind of laser undulator will open the path for many innovations.

ACKNOWLEDGMENTS
This work was supported by the National Natural Science Foundation of China (Grants No. 61690223, No. 11561121002, No. 61521093, No. 11127901, No. 11227902, No. 11404356, No. 11574332), the Strategic Priority Research Program of the Chinese Academy of Sciences (Grant No. XDB16), Shanghai Commission of Science and Technology Sailing Project (Grant No. 14YF1406000), Shanghai Institute of Optics and Fine Mechanics Specialized Research Fund (Grant No. 1401561J00) and Youth Innovation Promotion Association CAS. Corresponding author: zhinan_zeng@mail.siom.ac.cn; ruxinli@mail.shcnc.ac.cn;

[1] Brian W. J. McNeil and Neil R. Thompson, X-ray free-electron lasers, Nature Photon. 4, 814 (2010)
[2] Matthias Fuchs, Raphael Weingartner, Antonia Popp, Zsuzsanna Major, Stefan Becker, Jens Osterhoff, Isabella Cortrie, Benno Zeitler, Rainer Hörlein, George D. Tsakiris, Ulrich Schramm, Tom P. Rowlands-Rees, Simon M. Hooker, Dietrich Habs, Ferenc Krausz, Stefan Karsch and Florian Grüner, Laser-driven soft-X-ray undulator source, Nature Physics 5, 826 (2009)
[3] J. S. Liu, C. Q. Xia, W. T. Wang, H. Y. Lu, Ch. Wang, A. H. Deng, W. T. Li, H. Zhang, X. Y. Liang, Y. X. Leng, X. M. Lu, C. Wang, J. Z. Wang, K. Nakajima, R. X. Li, and Z. Z. Xu,

All-Optical Cascaded Laser Wakefield Accelerator Using Ionization-Induced Injection, Phys. Rev. Lett. 107, 035001 (2011)

[4] J. Faure, C. Rechatin, A. Norlin, A. Lifschitz, Y. Glinec & V. Malka, Controlled injection and acceleration of electrons in plasma wakefields by colliding laser pulses, Nature 444, 737 (2006)

[5] Xiaoming Wang, Rafal Zgadzaj, Neil Fazel, Zhengyan Li, S. A. Yi, Xi Zhang, Watson Henderson, Y.-Y. Chang, R. Korzekwa, H. -E. Tsai, C. -H. Pai, H. Quevedo, G. Dyer, E. Gaul, M. Martinez, A. C. Bernstein, T. Borger, M. Spinks, M. Donovan, V. Khudik, G. Shvets, T. Ditmire, and M. C. Downer, Quasi-monoenergetic laser-plasma acceleration of electrons to 2 GeV, Nature Comms 4, 1988 (2013)

[6] S. Steinke, J. van Tilborg, C. Benedetti, C. G. R. Geddes, C. B. Schroeder, J. Daniels, K. K. Swanson, A. J. Gonsalves, K. Nakamura, N. H. Matlis, B. H. Shaw, E. Esarey, and W. P. Leemans, Multistage coupling of independent laser-plasma accelerators, Nature 530, 190 (2016)

[7] Wentao Wang, Wentao Li, Jiansheng Liu, Cheng Wang, Qiang Chen, Zhijun Zhang, Rong Qi, Yuxin Leng, Xiaoyan Liang, Yanqi Liu, Xiaoming Lu, Cheng Wang, Ruxin Li, and Zhizhan Xu, Control of seeding phase for a cascaded laser wakefield accelerator with gradient injection, Appl. Phys. Lett. 103, 243501 (2013)

[8] S. M. Hooker, Developments in laser-driven plasma accelerators, Nature Photonics 7, 775 (2013)

[9] A. J. Gonsalves, K. Nakamura, C. Lin, D. Panasenko, S. Shiraishi, T. Sokollik, C. Benedetti, C. B. Schroeder, C. G. R. Geddes, J. van Tilborg, J. Osterhoff, E. Esarey, C. Toth and W. P. Leemans, Tunable laser plasma accelerator based on longitudinal density tailoring, Nature Physics 7, 862 (2011)

[10] Hyung Taek Kim, Ki Hong Pae, Hyuk Jin Cha, I Jong Kim, Tae Jun Yu, Jae Hee Sung, Seong Ku Lee, Tae Moon Jeong, and Jongmin Lee, Enhancement of Electron Energy to the Multi-GeV Regime by a Dual-Stage Laser-Wakefield Accelerator Pumped by Petawatt Laser Pulses, Phys. Rev. Lett. 111 165002 (2013)

[11] W. P. Wang, B. F. Shen, X. M. Zhang, X. F. Wang, J. C. Xu, X. Y. Zhao, Y. H. Yu, L. Q. Yi, Y. Shi, L. G. Zhang, T. J. Xu, and Z. Z. Xu, Cascaded target normal sheath acceleration, Phys. Plasmas 20, 113107 (2013)

[12] H. -P. Schlenvoigt, K. Haupt, A. Debus, F. Budde, O. Jäckel, S. Pfotenhauer, H. Schwoerer, E. Rohwer, J. G. Gallacher, E. Brunetti, R. P. Shanks, S. M. Wiggins and D. A. Jaroszynski, A compact synchrotron radiation source driven by a laser-plasma wakefield accelerator, Nature physics 4, 130 (2008)

[13] A. R. Maier, A. Meseck, S. Reiche, C. B. Schroeder, T. Seggebrock, and F. Grüner, Demonstration Scheme for a Laser-Plasma-Driven Free-Electron Laser, Phys. Rev. X 2, 031019 (2012)

[14] E. Esarey, C. B. Schroeder, and W. P. Leemans, Physics of laser-driven plasma-based electron accelerators, Reviews of Modern Physics 81, 1229 (2009)

[15] M E Couprie, A Loulergue, M Labat, R Lehe and V Malka, Towards a free electron laser based on laser plasma accelerators, J. Phys. B: At. Mol. Opt. Phys. 47, 234001 (2014)

[16] Zhirong Huang, Yuantao Ding, and Carl B. Schroeder, Compact X-ray Free-Electron Laser from a Laser-Plasma Accelerator Using a Transverse-Gradient Undulator, Phys. Rev. Lett. 109, 204801 (2012)

[17] F. Toufexis, T. Tang, S.G. Tantawi, A 200 μm-period laser-driven undulator, Proceedings of

FEL2014, Basel, Switzerland

[18] I. A. Andriyash, R. Lehe, A. Lifschitz, C. Thaury, J.-M. Rax, K. Krushelnick & V. Malka, An ultracompact X-ray source based on a laser-plasma undulator, Nature Communications 5, 4736 (2014)

[19] T. Plettner and R. L. Byer, Proposed dielectric-based microstructure laser-driven undulator, Physical review special topics – accelerators and beams 11, 030704 (2008)

[20] S. G. Rykovanov, C. B. Schroeder, E. Esarey, C. G. R. Geddes, and W. P. Leemans, Plasma Undulator Based on Laser Excitation of Wakefields in a Plasma Channel, Phys. Rev. Lett. 114, 145003 (2015)

[21] Chao Chang, Chuanxiang Tang, and Juhao Wu, High-Gain Thompson-Scattering X-Ray Free-Electron Laser by Time-Synchronic Laterally Tilted Optical Wave, Phys. Rev. Lett. 110, 064802 (2013)

[22] J E Lawler, J Bisognano, R A Bosch, T C Chiang, M A Green, K Jacobs, T Miller, R Wehlitz, D Yavuz and R C York, Nearly copropagating sheared laser pulse FEL undulator for soft x-rays, J. Phys. D: Appl. Phys. 46, 325501 (2013)

[23] K. Ta Phuoc, S. Corde, C. Thaury, V. Malka, A. Tafzi, J. P. Goddet, R. C. Shah, S. Sebban and A. Rousse, All-optical Compton gamma-ray source, Nature Photonics 6, 308 (2012)

[24] V. Petrillo, L. Serafini, and P. Tomassini, Ultrahigh brightness electron beams by plasma-based injectors for driving all-optical free-electron lasers, Phys. Rev. Special Topics - Accelerators and Beams 11, 070703 (2008)

[25] P. Sprangle, B. Hafizi, and J. R. Peñano, Laser-pumped coherent x-ray free-electron laser, Phys. Rev. Special Topics - Accelerators and Beams 12, 050702 (2009)

[26] Silvia Cipiccia, Mohammad R. Islam, Bernhard Ersfeld, Richard P. Shanks, Enrico Brunetti, Gregory Vieux, Xue Yang, Riju C. Issac, Samuel M. Wiggins, Gregor H. Welsh, Maria-Pia Anania, Dzmitry Maneuski, Rachel Montgomery, Gary Smith, Matthias Hoek, David J. Hamilton, Nuno R. C. Lemos, Dan Symes, Pattathil P. Rajeev, Val O. Shea, João M. Dias and Dino A. Jaroszynski, Gamma-rays from harmonically resonant betatron oscillations in a plasma wake, Nature Physics 7, 867–871 (2011)

[27] Chao Chang, Jinyang Liang, Dongwei Hei, Michael F. Becker, Kelei Tang, Yiping Feng, Vitaly Yakimenko, Claudio Pellegrini, and Juhao Wu, High-brightness X-ray free-electron laser with an optical undulator by pulse shaping, Opt. Express 21, 32013 (2013)

[28] Tong Zhang, Chao Feng, Haixiao Deng, Dong Wang, Zhimin Dai, and Zhentang Zhao, Compensating the electron beam energy spread by the natural transverse gradient of laser undulator in all-optical x-ray light sources, Opt. Express 22, 13880 (2014)

[29] K. Steiniger, R. Widera, R. Pausch, A. Debus, M. Bussmann, U. Schramm, Wave optical description of the Traveling-Wave Thomson-Scattering optical undulator field and its application to the TWTS-FEL, Nuclear Instruments and Methods in Physics Research A 740, 147–152 (2014)

[30] Min Chen, Ji Luo, Fei-Yu Li, Feng Liu, Zheng-Ming Sheng, and Jie Zhang, Tunable synchrotron-like radiation from centimeter scale plasma channels, Light Sci Appl 5 e16015 (2016)

[31] Brice Quesnel and Patrick Mora, Theory and simulation of the interaction of ultraintense laser pulses with electrons in vacuum, Phys. Rev. E 58, 3719 (1998)

[32] Ph. Balcou, Proposal for a Raman X-ray free electron laser, Eur. Phys. J. D 59, 525–537 (2010); I. A. Andriyasha, Ph. Balcou, and V.T. Tikhonchuk, Collective properties of a relativistic electron beam injected into a high intensity optical lattice, Eur. Phys. J. D 65, 533–540 (2011); I. A.

Andriyash, E. d'Humières, V. T. Tikhonchuk, and Ph. Balcou, X-Ray Amplification from a Raman Free-Electron Laser, Phys. Rev. Lett. 109, 244802 (2012); I. A. Andriyash, E. d'Humières, V. T. Tikhonchuk, Ph. Balcou, X-ray emission from relativistic electrons in a transverse high intensity optical lattice, Journal of Physics: Conference Series **414,** 012008(2013);